\documentclass[aps,12pt,superscriptaddress,amsfonts,amssymb,amsmath]{revtex4-2}
\pdfoutput=1

\usepackage{graphicx}
\usepackage{epsfig}
\usepackage{makeidx}
\usepackage{xcolor}
\usepackage{amsmath}
\usepackage{bm}
\usepackage{amssymb}
\usepackage[a4paper,margin=2.6cm]{geometry}
\usepackage{amsmath,amssymb}
\usepackage{hyperref}
\usepackage{float}
\usepackage{amsmath,amssymb}
\usepackage{pgfplots}
\usepgfplotslibrary{groupplots}
\pgfplotsset{compat=1.18}

\begin{document}

\title{
A Structural Link Between the Bohm Quantum Potential and the Scalar Mode of Aharonov--Bohm Electrodynamics in a Bosonic Schrödinger Model
} 

\author{R.\ Pullano \footnote{Email address: rosario.pullano@student.unibz.it}}
\affiliation{Free University of Bozen-Bolzano \\ Faculty of Engineering \\ I-39100 Bolzano, Italy}

\author{G.\ Modanese \footnote{Email address: giovanni.modanese@unibz.it}}
\affiliation{Free University of Bozen-Bolzano \\ Faculty of Engineering \\ I-39100 Bolzano, Italy}
\date{\today}

\linespread{0.9}

\begin{abstract}

We discuss a formal and physical connection between the Bohm quantum potential
and the scalar mode of the Aharonov--Bohm extension of electrodynamics. The
analysis is motivated by the effective non-relativistic bosonic model recently
proposed by Minotti and Modanese, in which the electromagnetic field is coupled
to a conserved current while the field equations contain an additional source
term. In the Madelung representation
$\psi=R\exp(i\theta/\hbar)$, the Bohm quantum potential
\(
Q_B=-\frac{\hbar^2}{2m}\frac{\nabla^2 R}{R}
\)
is determined by the relative curvature $\nabla^2R/R$ of the amplitude profile
$R$. In the same bosonic model, the scalar electromagnetic mode
$S=\partial_\mu A^\mu$ is sourced by the extra-current
$I=\partial_\mu j^\mu$, which contains the density-weighted electromagnetic
combination $\nabla\cdot(R^2\mathbf A)$. Thus $Q_B$ does not act as a direct
source of $S$; rather, the two quantities probe different differential aspects of
the same amplitude profile: $Q_B$ is sensitive to the relative curvature of $R$,
whereas the source of $S$ is sensitive to its density and gradient content through
$R^2$ and $\nabla R$. We show that, once boundary and normalization data are
fixed, this observation may be written as a mediated functional dependence of
$S$ on $Q_B$ through $R$. We also clarify the physical status of $Q_B$: although
it is state-dependent and should not be interpreted as an autonomous external
potential, its density-weighted integral gives the amplitude-gradient energy,
equivalently a Fisher-information contribution. This makes $Q_B$ a compact
diagnostic of quantum pressure, rigidity, and inhomogeneity of a bosonic
condensate. The resulting link with $S$ is therefore best understood as a
structural relation between the order-parameter amplitude profile of the
condensate and the scalar sector of the extended electromagnetic theory.
\end{abstract}

\maketitle

\section{Introduction}

The Aharonov--Bohm extension of electrodynamics promotes the four-divergence of the electromagnetic potential,
\begin{equation}
S \equiv \partial_\mu A^\mu ,
\end{equation}
to a dynamical scalar mode. In this framework the field equations can be coupled to sources that are not locally conserved in the ordinary Maxwell sense. If $j^\mu$ is the current entering the electromagnetic equations, the extra-current
\begin{equation}
I\equiv \partial_\mu j^\mu
\end{equation}
acts as a source for $S$. In the normalization used below,
\begin{equation}
\Box S = \mu_0 I .
\end{equation}
This structure has recently acquired renewed interest in effective descriptions of bosonic matter, in particular in models where a macroscopic wavefunction describes charged scalar quasiparticles such as Cooper pairs. Minotti and Modanese have argued that, for scalar bosonic systems, an alternative coupling to the extended Aharonov--Bohm electrodynamics leads, in the non-relativistic limit, to a Schrödinger-type equation containing an additional term proportional to $\mathbf A^2$ and to an electromagnetic source containing a contribution proportional to $|\psi|^2\mathbf A$ \cite{MinottiModanese2026}.

The purpose of this note is to point out that this bosonic structure naturally brings the scalar Aharonov--Bohm mode into contact with the Bohm quantum potential. We use the Madelung decomposition
\begin{equation}
\psi(\mathbf x,t)=R(\mathbf x,t)e^{i\theta(\mathbf x,t)/\hbar}, \qquad R\geq 0,
\end{equation}
where $R^2$ is the density associated with the macroscopic quantum state. In the hydrodynamic form of the Schrödinger equation, the real part of the dynamics contains the Bohm quantum potential
\begin{equation}
Q_B(\mathbf x,t)=-\frac{\hbar^2}{2m}\frac{\nabla^2R}{R}.
\end{equation}
The same amplitude $R$ also enters the extra-current that sources the scalar mode $S$ in the bosonic Aharonov--Bohm model. More precisely, the relevant source contains
\begin{equation}
\nabla\cdot(R^2\mathbf A),
\end{equation}
so that the electromagnetic scalar sector is sensitive to the same density amplitude inhomogeneity that generates $Q_B$.

This observation must be formulated with some care. The Bohm potential is not an external potential like the Coulomb potential of a fixed charge distribution. It is a state-dependent functional of the amplitude of the wavefunction or order parameter. Therefore, the relation to $S$ should not be read as a direct causal identification $Q_B\to S$. The physically meaningful statement is more structural: in a bosonic Schrödinger-like system coupled to Aharonov--Bohm electrodynamics, the scalar field $S$ is sourced by the same amplitude $R(\mathbf x,t)$ that generates the Bohm quantum potential. Consequently, $Q_B$ can be used as a compact diagnostic of the condensate inhomogeneity relevant for the scalar electromagnetic sector. The relation is mediated by the order-parameter amplitude $R$, not by an independent force field represented by $Q_B$ alone.

This viewpoint also clarifies why the connection may have physical content despite the state-dependence of $Q_B$. The local quantity $Q_B(\mathbf x,t)$ depends on the particular quantum state, but its density-weighted integral is directly related, up to boundary terms, to the gradient energy of the amplitude:
\begin{equation}
\int R^2 Q_B\,d^3x
= -\frac{\hbar^2}{2m}\int R\nabla^2R\,d^3x
= \frac{\hbar^2}{2m}\int |\nabla R|^2\,d^3x
\end{equation}
for boundary conditions that make the surface term vanish. In condensates and superfluids this term is commonly interpreted as quantum pressure, healing energy, interface energy, or the energetic cost of spatial inhomogeneity. Thus $Q_B$ is not an autonomous material potential, but it is the local dynamical representative of a physically meaningful texture energy. This is analogous, in a limited but useful sense, to the internal energy of a macroscopic system: its microscopic value depends on the state, while its thermodynamic role becomes operational when expressed in terms of suitable collective variables.

A complementary motivation comes from the broader geometrical and informational readings of the quantum potential. In the survey by Licata and Fiscaletti, the quantum potential is not treated merely as a formal correction to the classical Hamilton--Jacobi equation, but as a central object in several physical, geometrical, algebraic, and informational reconstructions of quantum theory \cite{LicataFiscaletti2014}. In particular, the quantum potential is often associated with active information, nonlocal constraints, the geometry of configuration space, and Fisher-information-like measures of the amplitude distribution. In the present problem, this is especially suggestive: the link between $Q_B$ and $S$ does not pass through an ordinary mechanical force, but through the amplitude/density structure of a macroscopic quantum state. The same amplitude profile is read in one sector as quantum pressure or information geometry and in the other as an extra-current sourcing the scalar electromagnetic mode.

The paper is organized as follows. Section \ref{sec:standard} recalls the Madelung form of the minimally coupled Schrödinger equation and the definition of $Q_B$. Section \ref{sec:bosonic} applies the same decomposition to the bosonic model with the additional $\mathbf A^2$ term and identifies the effective current that sources the Aharonov--Bohm scalar mode. Section \ref{sec:mediated} formulates the mediated functional relation between $S$ and $Q_B$. Section \ref{sec:physical} discusses the physical status of the Bohm potential, its relation to gradient energy and Fisher information, and its geometrical interpretation. Section \ref{sec:examples} gives simple reductions that make the relation explicit. The conclusions summarize the structural content and the limitations of the proposed link.

\section{Bohm potential from the minimally coupled Schrödinger equation}
\label{sec:standard}

Let
\begin{equation}
A^\mu = (\phi/c,\mathbf A)
\end{equation}
be the electromagnetic four-potential. We first recall the standard Schrödinger equation with electromagnetic potentials and an external scalar potential $V$:
\begin{equation}
 i\hbar\,\partial_t\psi=
\left[\frac{1}{2m}(-i\hbar\nabla-q\mathbf A)^2+q\phi+V\right]\psi .
\label{standardSch}
\end{equation}
With
\begin{equation}
\psi=R e^{i\theta/\hbar}
\end{equation}
and
\begin{equation}
\mathbf u\equiv \nabla\theta-q\mathbf A,
\end{equation}
the imaginary part of Eq.~\eqref{standardSch} gives the continuity equation
\begin{equation}
\partial_t(R^2)+\nabla\cdot\left(\frac{R^2}{m}\mathbf u\right)=0 .
\end{equation}
If
\begin{equation}
\rho=qR^2,
\qquad
\mathbf j=\frac{q}{m}R^2(\nabla\theta-q\mathbf A),
\end{equation}
then
\begin{equation}
\partial_t\rho+\nabla\cdot\mathbf j=0.
\end{equation}
The real part gives the quantum Hamilton--Jacobi equation
\begin{equation}
-\partial_t\theta=
\frac{(\nabla\theta-q\mathbf A)^2}{2m}+q\phi+V+Q_B,
\end{equation}
where
\begin{equation}
Q_B(\mathbf x,t)=-\frac{\hbar^2}{2m}\frac{\nabla^2R}{R}.
\label{QBdef}
\end{equation}
The Bohm potential enters the Hamilton--Jacobi equation in the same energy balance as the kinetic, electromagnetic, and external potential terms. Its origin, however, is different: it is generated by the curvature of the amplitude of the quantum state.

\section{Bosonic Schrödinger model and Aharonov--Bohm scalar source}
\label{sec:bosonic}

In the effective bosonic model considered here, the non-relativistic equation contains an additional term proportional to $\mathbf A^2$:
\begin{equation}
 i\hbar\,\partial_t\psi=
\left[\frac{1}{2m}(-i\hbar\nabla-q\mathbf A)^2+q\phi+V+\frac{q^2}{2m}\mathbf A^2\right]\psi .
\label{bosonicSch}
\end{equation}
The charge density is assumed to be
\begin{equation}
\rho=q|\psi|^2=qR^2.
\end{equation}
The conserved current associated with the matter sector is
\begin{equation}
\mathbf j_c=\frac{iq\hbar}{2m}(\psi\nabla\psi^*-\psi^*\nabla\psi)-\frac{q^2}{m}|\psi|^2\mathbf A .
\label{jcdef}
\end{equation}
Using the Madelung form of $\psi$, one obtains
\begin{equation}
\mathbf j_c=\frac{q}{m}R^2\nabla\theta-\frac{q^2}{m}R^2\mathbf A
=\frac{q}{m}R^2(\nabla\theta-q\mathbf A).
\end{equation}
Thus the conserved current keeps the standard hydrodynamic structure, and
\begin{equation}
\partial_t\rho+\nabla\cdot\mathbf j_c=0.
\label{conserved}
\end{equation}
The real part of Eq.~\eqref{bosonicSch} gives
\begin{equation}
-\partial_t\theta=
\frac{(\nabla\theta-q\mathbf A)^2}{2m}+q\phi+V+Q_B+\frac{q^2}{2m}\mathbf A^2 .
\label{HJbosonic}
\end{equation}
The Bohm potential is the same as in Eq.~\eqref{QBdef}; the additional $\mathbf A^2$ contribution modifies the total Hamilton--Jacobi balance but not the definition of $Q_B$.

In the corresponding Aharonov--Bohm electrodynamics, the current sourcing the field contains, besides $\mathbf j_c$, an additional term. We define the effective field current
\begin{equation}
\mathbf J_{\rm field}\equiv \mathbf j_c-\frac{q^2}{m}R^2\mathbf A .
\label{Jfield}
\end{equation}
The extra-current that sources $S$ is then
\begin{equation}
I\equiv \partial_t\rho+\nabla\cdot\mathbf J_{\rm field}.
\end{equation}
Using Eq.~\eqref{conserved}, one obtains
\begin{equation}
I=-\frac{q^2}{m}\nabla\cdot(R^2\mathbf A).
\label{Ibasic}
\end{equation}
Equivalently,
\begin{equation}
I=-\frac{q^2}{m}\left(R^2\nabla\cdot\mathbf A+2R\mathbf A\cdot\nabla R\right).
\label{Iexpanded}
\end{equation}
For $\lambda=1$, the scalar Aharonov--Bohm mode therefore satisfies
\begin{equation}
\Box S=\mu_0I
=-\mu_0\frac{q^2}{m}\nabla\cdot(R^2\mathbf A),
\label{Ssource}
\end{equation}
or, explicitly,
\begin{equation}
\Box S=-\mu_0\frac{q^2}{m}\left(R^2\nabla\cdot\mathbf A+2R\mathbf A\cdot\nabla R\right).
\label{SsourceExpanded}
\end{equation}
Equations \eqref{Ibasic}--\eqref{SsourceExpanded} are the central observation: the scalar electromagnetic mode is sourced by the density and gradient of the same amplitude $R$ that generates the Bohm quantum potential.

\section{Mediated dependence of $S$ on $Q_B$ through the amplitude}
\label{sec:mediated}

From Eq.~\eqref{QBdef}, one obtains
\begin{equation}
Q_B=-\frac{\hbar^2}{2m}\frac{\nabla^2R}{R}
\quad\Longleftrightarrow\quad
\nabla^2R+\frac{2m}{\hbar^2}Q_B R=0 .
\label{RfromQ}
\end{equation}
Thus, after boundary conditions and normalization are specified, $R$ may be regarded formally as a functional of $Q_B$:
\begin{equation}
R=R[Q_B].
\end{equation}
Substitution into Eq.~\eqref{Ssource} gives
\begin{equation}
\Box S=-\mu_0\frac{q^2}{m}\nabla\cdot\left(R[Q_B]^2\mathbf A\right).
\label{functionalRelation}
\end{equation}
With a Green operator $\Box^{-1}$ chosen consistently with the boundary or radiation conditions, one may also write
\begin{equation}
S=-\mu_0\frac{q^2}{m}\Box^{-1}\left[\nabla\cdot\left(R[Q_B]^2\mathbf A\right)\right].
\label{GreenS}
\end{equation}

Equations \eqref{functionalRelation} and \eqref{GreenS} express a formal mediated dependence of $S$ on $Q_B$. This dependence is not one-to-one in a purely local sense. In particular, $Q_B$ is invariant under a constant rescaling $R\mapsto cR$, while the source $\nabla\cdot(R^2\mathbf A)$ is not. The scale of the density and the boundary data are therefore essential. Moreover, the definition of $Q_B$ requires care at nodes of $R$. For these reasons the physical claim should be stated as follows: $Q_B$ and $S$ are structurally linked because both depend on the same amplitude. The Bohm potential measures the relative curvature of that amplitude, whereas the Aharonov--Bohm scalar source measures how the density couples to $\mathbf A$ through $\nabla\cdot(R^2\mathbf A)$.

It is useful to introduce
\begin{equation}
W\equiv \ln R,\qquad R>0.
\end{equation}
Then
\begin{equation}
\frac{\nabla^2R}{R}=\nabla^2W+|\nabla W|^2,
\end{equation}
and
\begin{equation}
Q_B=-\frac{\hbar^2}{2m}\left(\nabla^2W+|\nabla W|^2\right).
\label{QW}
\end{equation}
The scalar source becomes
\begin{equation}
\Box S=-\mu_0\frac{q^2}{m}\nabla\cdot(e^{2W}\mathbf A)
=-\mu_0\frac{q^2}{m}e^{2W}\left(\nabla\cdot\mathbf A+2\mathbf A\cdot\nabla W\right).
\label{SW}
\end{equation}
This form shows why the relation cannot be closed by replacing $\nabla^2W$ through $Q_B$ alone: the scalar source also depends on the first-derivative geometry $\nabla W$ and on the absolute density factor $e^{2W}$.

\section{Physical status of the Bohm potential: texture energy, information, and geometry}
\label{sec:physical}

The state-dependence of $Q_B$ is sometimes viewed as an obstacle to assigning it physical significance. In the present context, however, it is precisely this state-dependence that makes $Q_B$ relevant: it is a local descriptor of the spatial organization of the condensate amplitude. A non-uniform $R(\mathbf x,t)$ describes domains, interfaces, healing layers, or other inhomogeneous structures of the bosonic order parameter. The Bohm potential is the Hamilton--Jacobi representative of the relative curvature of this amplitude.

A more operational quantity is obtained by weighting $Q_B$ with the density. Assuming that the surface contribution vanishes, integration by parts gives
\begin{equation}
\int R^2 Q_B\,d^3x
= -\frac{\hbar^2}{2m}\int R\nabla^2R\,d^3x
= \frac{\hbar^2}{2m}\int |\nabla R|^2\,d^3x .
\label{gradEnergy}
\end{equation}
The right-hand side is the amplitude-gradient energy. In hydrodynamic and condensate language, it is naturally associated with quantum pressure, stiffness, and the energetic cost of spatial inhomogeneity. This provides a physical interpretation that is weaker than treating $Q_B$ as an autonomous external field, but stronger than treating it as a merely formal rewriting of the Schrödinger equation.

The same expression has an informational reading. For a normalized density $\rho_R=R^2$, the Fisher information is
\begin{equation}
\mathcal I_F[\rho_R]=\int \frac{|\nabla\rho_R|^2}{\rho_R}\,d^3x
=4\int |\nabla R|^2\,d^3x .
\end{equation}
Therefore Eq.~\eqref{gradEnergy} may be written as
\begin{equation}
\int R^2 Q_B\,d^3x=\frac{\hbar^2}{8m}\mathcal I_F[\rho_R].
\label{FisherRelation}
\end{equation}
This identity makes explicit that the integrated quantum-potential contribution is a measure of the sharpness or spatial information content of the amplitude distribution. It also provides a natural bridge to the geometrical and informational interpretations of the quantum potential discussed in the literature \cite{LicataFiscaletti2014,Frieden2004,Hall2000,FiscalettiLicata2012}.

In the language of Licata and Fiscaletti, the quantum potential is a flexible object that can be read dynamically, geometrically, thermodynamically, algebraically, and informationally. The common thread is that $Q_B$ encodes the way in which the form of the quantum state constrains the motion or field configuration. In the present application, this interpretation is particularly apt. The scalar mode $S$ is not sourced by a classical charge imbalance alone, but by the failure of the effective field current to be locally conserved, and that failure is controlled by $R^2\mathbf A$. Thus the scalar electromagnetic sector is sensitive to the same amplitude geometry whose curvature appears as $Q_B$ in the Hamilton--Jacobi sector.

This allows the following physical picture. The order-parameter amplitude $R$ is the primary texture. Its curvature gives the local Bohm potential, its gradient energy gives a measurable stiffness or quantum-pressure contribution, its Fisher information measures the spatial sharpness of the density, and its electromagnetic weighting $\nabla\cdot(R^2\mathbf A)$ gives the extra-current that sources $S$. In this sense, $Q_B$ is not the source of $S$ by itself. Rather, $Q_B$ is a compact diagnostic of the same texture of the bosonic condensate that, in the Aharonov--Bohm extension, becomes electromagnetically active through the scalar mode.

\section{Simple reductions}
\label{sec:examples}

\subsection{One-dimensional effective case}

In one spatial dimension, with $W=W(x)$, Eq.~\eqref{QW} gives
\begin{equation}
W''(x)=-\frac{2m}{\hbar^2}Q_B(x)-[W'(x)]^2.
\label{Riccati}
\end{equation}
This is a Riccati-type equation for $W'(x)$, once $Q_B(x)$ is specified. If
\begin{equation}
\mathbf A=A_x(x)\hat{\mathbf x},
\end{equation}
then
\begin{equation}
\nabla\cdot(e^{2W}\mathbf A)=\partial_x(e^{2W}A_x)
=e^{2W}(A_x'+2A_xW'),
\end{equation}
and
\begin{equation}
\Box S=-\mu_0\frac{q^2}{m}\partial_x(e^{2W}A_x).
\label{S1D}
\end{equation}
In this simplified setting the mediated chain
\begin{equation}
Q_B \longrightarrow W \longrightarrow R=e^W \longrightarrow I \longrightarrow S
\end{equation}
is computationally explicit, once the required boundary and normalization data are provided.

\subsection{Geometric ansatz}

One may also impose a geometric ansatz relating the amplitude gradient to the vector potential, for example
\begin{equation}
\nabla W=\kappa(\mathbf x,t)\mathbf A.
\end{equation}
Then
\begin{equation}
\mathbf A\cdot\nabla W=\kappa\mathbf A^2,
\qquad
|\nabla W|^2=\kappa^2\mathbf A^2,
\end{equation}
and Eq.~\eqref{SW} becomes
\begin{equation}
\Box S=-\mu_0\frac{q^2}{m}e^{2W}
\left(\nabla\cdot\mathbf A+2\kappa\mathbf A^2\right).
\end{equation}
This type of ansatz reduces the number of independent geometric degrees of freedom and makes the projection of the amplitude inhomogeneity along the electromagnetic potential explicit.

\section{Conclusions}
\label{sec:conclusions}

We have shown that, in the effective bosonic Schrödinger model coupled to Aharonov--Bohm extended electrodynamics, the scalar mode
\begin{equation}
S=\partial_\mu A^\mu
\end{equation}
is sourced by the same amplitude inhomogeneity that generates the Bohm quantum potential. The Madelung decomposition gives
\begin{equation}
Q_B=-\frac{\hbar^2}{2m}\frac{\nabla^2R}{R},
\end{equation}
while the extra-current entering the Aharonov--Bohm scalar equation is
\begin{equation}
I=-\frac{q^2}{m}\nabla\cdot(R^2\mathbf A).
\end{equation}
Consequently,
\begin{equation}
\Box S=-\mu_0\frac{q^2}{m}\nabla\cdot(R^2\mathbf A).
\end{equation}
After boundary and normalization data are fixed, the equation defining $Q_B$ may be read formally as an equation for $R$, leading to
\begin{equation}
\Box S=-\mu_0\frac{q^2}{m}\nabla\cdot\left(R[Q_B]^2\mathbf A\right).
\end{equation}
This is the precise sense in which $S$ may be said to depend on $Q_B$: the dependence is mediated by the amplitude $R$.

The main physical point is not that $Q_B$ is an autonomous source field. The Bohm potential is a state-dependent functional of the amplitude. Nevertheless, this does not make it physically empty. Its density-weighted integral is the amplitude-gradient energy and, for a normalized density, is proportional to the Fisher information. In a bosonic condensate this quantity measures quantum pressure, stiffness, healing, and the cost of spatial inhomogeneity. Thus $Q_B$ has the status of a local diagnostic of a physically meaningful order-parameter texture.

The link with Aharonov--Bohm electrodynamics is therefore structural. The same inhomogeneous condensate profile is read in the Hamilton--Jacobi sector as the Bohm potential and in the extended electromagnetic sector as an extra-current sourcing $S$. This gives the proposed connection a clear interpretation: the scalar Aharonov--Bohm mode is sensitive to condensate textures whose quantum-pressure content is compactly encoded by $Q_B$.

The formulation also displays its own limitations. The map from $Q_B$ to $R$ is not unique without boundary conditions, normalization, and a treatment of nodes. Moreover, $Q_B$ is invariant under constant rescalings of $R$, whereas the scalar source depends on $R^2$. For this reason the relation should not be presented as a direct causal law $Q_B\to S$, but as a mediated relation through the macroscopic amplitude of the bosonic state. Within these qualifications, the result suggests a useful diagnostic bridge between Bohmian hydrodynamics, information-geometric readings of the quantum potential, and the scalar sector of Aharonov--Bohm electrodynamics.

\end{document}